%% file: blackfire_v2.tex
\documentclass[11pt,a4paper,svgnames]{article}
\pdfoutput=1

\usepackage[utf8]{inputenc}
\usepackage{uniinput}
\usepackage{jheppubmod}
\usepackage{hyperref}
\usepackage[plain]{fancyref}
\usepackage{amsmath}
\usepackage{url}
\usepackage{bm}
\usepackage{mathtools}
\usepackage{relsize}
\usepackage{cancel}
\usepackage{caption,subcaption}
\usepackage{lipsum}

\newcommand{\D}{\mathrm{d}}
\newcommand{\ie}{\textit{i.e.}}
\newcommand{\eg}{\textit{e.g.}}
\newcommand{\bea}{\begin{eqnarray}}
\newcommand{\eea}{\end{eqnarray}}
\newcommand{\be}{\begin{equation}}
\newcommand{\ee}{\end{equation}}
\usepackage[format=plain, textfont=it]{caption}
\usepackage{microtype}
\allowdisplaybreaks

\graphicspath{{figures/}}

\title{Dark bubbles and black holes}
\author[a]{Souvik Banerjee,}
\emailAdd{souvik.banerjee@physik.uni-wuerzburg.de}
\affiliation[a]{Institut für Theoretische Physik und Astrophysik,
	Julius-Maximilians-Universität Würzburg,Am Hubland, 97074 Würzburg, Germany\\}
\author[b]{Ulf Danielsson,}
\emailAdd{ulf.danielsson@physics.uu.se}
\affiliation[b]{Institutionen för fysik och astronomi,
	Uppsala Universitet, Box 803, SE-751 08 Uppsala, Sweden\\}
\author[c,d]{Suvendu Giri~}
\emailAdd{suvendu.giri@unimib.it}
\affiliation[c]{Dipartimento di Fisica, Università di Milano-Bicocca, I-20126 Milano, Italy}
\affiliation[d]{INFN, sezione di Milano-Bicocca, I-20126 Milano, Italy}

\abstract{
In this paper we study shells of matter and black holes on the expanding bubbles realizing de Sitter space, that were proposed in arXiv:1807.01570. We construct explicit solutions for a rigid shell of matter as well as black hole like solutions. The latter of these can also be used to construct Randall-Sundrum braneworld black holes in four dimensions.
}
\preprint{UUITP-05/21}

\begin{document}
\maketitle
\section{Introduction}\label{sec:intro}

To construct a model of de Sitter space and dark energy in string theory is a great challenge. Over the years, evidence has accumulated suggesting that many, possibly all, attempts made so far suffer from instabilities \cite{Danielsson:2018ztv,Obied:2018sgi}. For a review, see \cite{Palti:2019pca}. It is therefore of great importance to find alternative routes towards finding dS space. In \cite{Banerjee:2018qey}, we proposed that dark energy can be realized through an expanding bubble of true vacuum in a metastable AdS$_5$. Contrary to previous attempts, our focus is not on obtaining a time independent, metastable string vacuum with a positive vacuum energy. Instead, our model makes explicit use of an unstable higher dimensional AdS space with our universe riding on a bubble of true vacuum\footnote{we call this a \emph{dark bubble} since it is a bubble that gives rise to dark energy in four dimensions.} that mediates the decay of the unstable AdS.  Our model has much in common with the braneworlds of Randall and Sundrum, \cite{Randall:1999ee,Randall:1999vf}, but there are also crucial differences as explained in \fref{sec:review}.

Our model was further studied in \cite{Banerjee:2019fzz}, where we showed, using a five dimensional bulk with stretched strings, how this will, through junction conditions generate an effective theory of four dimensional Einstein gravity on our dark bubble. In subsequent works, \cite{Banerjee:2020wix, Banerjee:2020wov}, we worked out how to imprint a Schwarzschild geometry on the dark bubble and the back-reaction thereof on the bulk. For other aspects of these dark bubbles see, e.g., \cite{Koga:2019yzj,Basile:2020mpt,Dibitetto:2020csn,Rosu:2020tov}.

In this paper, we work out in detail, the embedding of certain four dimensional structures on the dark bubble into AdS$_5$ for two illuminating examples. We first examine a thin shell of matter, stabilized by its internal pressure, and then move on to consider a black hole. The new five dimensional metric that we find can also be used to realize black holes in four dimensional Randall-Sundrum (RS) braneworlds. A key observation in our paper is that the five dimensional spacetimes above (outside) and below (inside) our dark bubble are very different. We show how these two different views, one from above and the other from below, project the same effective four dimensional theory of gravity on the dark bubble in a rather miraculous way.

The rest of the paper is organized as follows. We begin with a brief review of the dark bubble model. Then in \fref{sec:rigid_shell}, we discuss the first example of a four dimensional rigid spherical shell of matter and demonstrate the uplifting of this construction to the five dimensional bulk spacetimes outside and inside of our dark bubble. While outside the dark bubble, we have a solution sourced by a stringy distribution, there is no source below and therefore, the junction condition across demands a non trivial choice of boundary conditions imposed on the brane from the bulk spacetime on either side. In \fref{sec:blackholes}, we move to our second example, namely, exploring the possibility of having a black hole on our bubble wall.\footnote{See \cite{Chamblin:1999by, Emparan:1999wa, Emparan:1999fd, Giddings:2000mu, Dadhich:2000am} for some of the earliest works towards constructing black holes on RS braneworlds. See \cite{Gregory:2008rf} and references therein, for a review of RS braneworld black holes and a more exhaustive list of work in this direction.} 
We show that it is possible in our construction by considering different non-linear corrections in gravitational perturbation theory. Finally, we conclude with an invitation to upcoming research activities in this direction.

\section{Brief review of dark bubbles} \label{sec:review}

Let us briefly summarize the most important features of our model. As a simple example we consider a spherical bubble of AdS$_5$ (with cosmological constant $Λ_-$) expanding in an AdS$_5$ spacetime  with a larger cosmological constant  $Λ_+>Λ_-$, where $\Lambda_\pm=-6k^2_\pm$. The surface of the bubble represents a Friedmann cosmology with the induced metric
\be
\D s²_4=-\D τ²+a(τ)²\D Ω²₃ \, .
\ee
With the AdS$_5$ metrics written in global coordinates $\D s^2_5 = -(1+k_\pm^2r^2)\D t^2 + \D r^2/(1+k_\pm^2r^2) + r^2 \D Ω²₃$, 
the Israel junction conditions force the radius of the bubble, $a(\tau )$, where $\tau$ is proper time, to obey
\be
\label{eq:junct}
\sigma=\frac{3}{8πG_5}\left(\sqrt{k^2_{-}+\frac{1+\dot{a}^2}{a^2}}-\sqrt{k_{+}^2+\frac{1+\dot{a}^2}{a^2}}\right)\,,
\ee
where $\sigma$ is the tension of the brane supporting the bubble wall.
For large $k_\pm$ we find
\be
\frac{\dot{a}²}{a²} \approx-\frac{1}{a²}+\frac{8πG₄}{3}Λ₄\,,
\ee
which is nothing else than the Friedmann equation (with the four dimensional Newton's constant obtained from the five dimensional one through $G_4=\frac{2k_- k_+ }{k_- -k_+} G_5$) in the presence of a positive cosmological constant given by $\Lambda_4 = \frac{3\left(k_{-}-k₊\right)}{8πG_5 }- \sigma$. One can show that the induced metric on the bubble wall is described by the full four dimensional Einstein equations \cite{Banerjee:2019fzz}.

One can generate radiation on the four dimensional world through a combination of radiation on the bubble wall and a nontrivial metric inside and outside of the bubble. In the presence of an AdS-Schwarzschild metric with ADM mass, $M_{\pm}$, the Friedmann equation becomes:
\be
\frac{\dot{a}²}{a²} \approx -\frac{1}{a²}+\frac{8π}{3}G₄\left(Λ₄+\frac{3}{4π²}\left(\frac{M₊}{k₊}-\frac{M_-}{k_-}\right)\frac{1}{a⁴}\right) .
\ee
We see, for instance, how $M_+>0$ on the outside but $M_-=0$ on the inside leads to a positive density of radiation. 

At this point it is useful to contrast our model to the braneworlds of Randall and Sundrum, \cite{Randall:1999ee,Randall:1999vf}. Contrary to those models, where the inside of the bubble is identified with itself across the brane, our bubble has an inside and an outside. This is why there is a minus sign between the two terms in the junction conditions, which carry over to the expression for $G_4$ as well as to the effective 4D radiation density. 

As was shown in \cite{Banerjee:2020wix}, you can get a massive particle if you let a string pull upwards from the brane. With a homogeneous distribution of many such strings, one reproduces the Friedmann equations in the presence of dust. What happens is that the dark bubble eats the strings as it expands. The energy from the strings is used so that the effective energy density on the bubble, and hence $H^2$, decays only as $1/a^3$ rather than $1/a^4$ as in the case of radiation. Let us now move on and consider some other interesting configurations.

\section{A rigid shell of matter uplifted in the fifth dimension}\label{sec:rigid_shell}

\begin{figure}[h]
    \centering
    \begin{subfigure}[b]{0.45\textwidth}
        \def\svgwidth{\linewidth}
        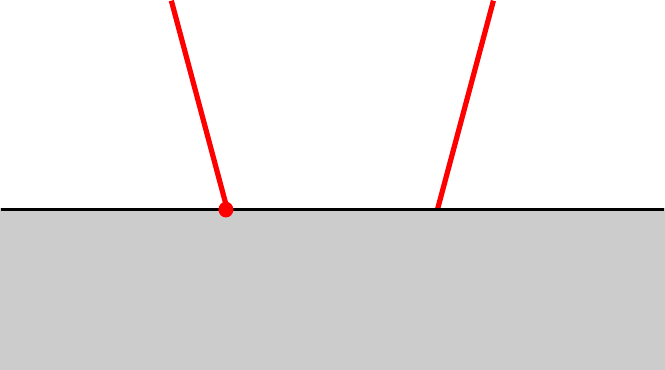  
        \caption{}   
    	\label{fig:structure_above_below:a}
    \end{subfigure}
    \hfill
    \begin{subfigure}[b]{0.45\textwidth}
        \def\svgwidth{\linewidth}
        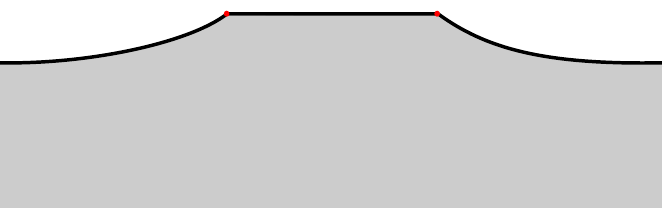
        \caption{}
		\label{fig:structure_above_below:b}
    \end{subfigure}
\label{fig:structure_above_below}
\caption{A cross-section of  a thin, rigid shell of four dimensional matter on the dark bubble seen \emph{(a)} from outside, and \emph{(b)} from inside. The two figures are given in two different gauges as explained in the text. The gray region indicates the inside of the dark bubble.}
\end{figure}

To better understand the physics of the four dimensional spacetime on the dark bubble, it is illustrative to consider a thin, rigid shell of matter with a radius $r_0$ much larger than its Schwarzschild radius. This could, for instance, be a shell made of ordinary baryonic matter such as iron. We showed in \cite{Banerjee:2018qey, Banerjee:2019fzz} that a point mass in four dimensions is the end point of a string located at a fixed radial distance, stretching in the fifth dimension. Following this intuition, the five dimensional structure corresponding to the rigid shell on the dark bubble would be expected to be located at a fixed coordinate radius and stretching in the fifth dimension. The proper radius $r_p \equiv krz$ then increases towards the boundary, giving a conical structure as sketched in \fref{fig:structure_above_below:a}.

The matter imprint on the dark bubble will then be interpreted as the  holographic projection of this five dimensional object into four dimensions. Just as a shell of matter in four dimensions has a complicated equation of state (so as to remain stable against collapse but still possible to deform), the same would be expected to be true for the five dimensional extended structure.

The five dimensional picture below the dark bubble is, however, quite different. There are no matter sources in this region, and the spacetime must be a solution of the Einstein's equations with just a negative cosmological constant. The existence, and form, of this solution is quite surprising. As we will show in the next section, the solution contains a holographic ``shadow'' of the structure above the dark bubble, and the resulting picture will look like \fref{fig:structure_above_below:b}. We will also see how the construction nicely generalizes the simpler examples studied in \cite{Banerjee:2020wix, Banerjee:2020wov}, where we made use of some techniques introduced in \cite{Giddings:2000mu,Padilla:2004mc}.
\subsection{Uplifting upwards}\label{sec:rigid_shell:up}
Let us first consider the five dimensional bulk spacetime above the brane as sketched in \fref{fig:structure_above_below:a}. Above the brane, it is convenient to work in a gauge in which the brane is flat. Spacetime outside the structure is expected to be that of a matter source. Since this structure lies far outside its own Schwarzschild radius, the metric just outside it should be given by the
CHR metric\footnote{This is the metric sourced by a neutral black string in five dimensions and induces four dimensional Schwarzschild geometry on constant $z$ slices. Symmetry arguments similar to Birkhoff's theorem would lead to the conclusion that the metric outside a cylindrically symmetric matter distribution is given by the CHR metric outside the four dimensional Schwarzschild radius of such a structure.} \cite{Chamblin:1999by}
\begin{equation}
    \D s² = k_+²z² \left[-\left(1-\frac{2M}{r}\right)\D t² + \frac{\D r²}{1-2M/r}+r²\D \Omega_2²\right] + \frac{\D z²}{k_+²z²},
\end{equation}
while the spacetime inside is empty AdS$_5$. For simplicity, we have taken the radius of the bubble (which we now call $z$ to be very large compared to the AdS radius $1/k_+$, so that the bubble wall is approximately flat and the metric looks like the Poincaré patch metric. The structure that supports the rigid shell of matter on the brane is located at $r=r_0=\text{constant}$. In terms of the proper time ($τ$) on this structure, the induced metric is simply $\D s² = k_+²z² \left(-\D τ² + r_0²\D \Omega_2²\right) + \D z²/k_+²z²$. The stress tensor on it ($S^i_j$) is determined by the thin-shell junction conditions matching five dimensional AdS to CHR and reads
\begin{equation}\label{eq:stresstensor4d}
\begin{aligned}
    S^τ_τ &= -\frac{2}{k_+ r_0 z}\left(1-\sqrt{1-\frac{2M}{r_0}}\right),\\
    S^θ_θ = S^ϕ_ϕ &= \frac{1}{k_+ r_0 z}\left(\frac{1-M/r_0}{\sqrt{1-2M/r_0}}-1\right),\\
    S^z_z &= \frac{2}{k_+ r_0 z}\left(\frac{1-3M/2r_0}{\sqrt{1-2M/r_0}}-1\right).
\end{aligned}
\end{equation}
When this structure ends on the dark bubble, Bianchi identities, which require the stress tensor to be covariantly conserved, produce a delta function contribution on the dark bubble at $z=z_0$. To see this explicitly, let us write down the covariant conservation equation $\nabla_{μ}T^{μν}\overset{!}{=}0$ for $ν=z$ ($T^\mu_\nu$ refers to the five dimensional stress tensor). Explicitly, this gives
\begin{equation}\label{eq:covd}
\begin{aligned}
    \nabla_{μ}T^{μν} &= ∂_μ T^{μν} + Γ^σ_{γσ} T^{γν} + Γ^ν_{σμ}T^{σμ} \overset{!}{=} 0\\
    &\Rightarrow z ∂_z T^z_z + 4 T^z_z \overset{!}{=} T^t_t + T^r_r + T^θ_θ + T^ϕ_ϕ.
\end{aligned}
\end{equation}
The stress tensor obtained in \fref{eq:stresstensor4d}, dressed up with a properly normalized delta function\footnote{the properly normalized delta function is $δ(r-r_0)/\sqrt{g_{rr}} = δ(r-r_0)\sqrt{1-2M/r}/kz$, so that it integrates to unity along the radial direction \ie, $\int \sqrt{g_{rr}} δ(r-r_0)/\sqrt{g_{rr}} = 1$.} localized at $r=r_0$ gives
\begin{equation}
    T^i_j = S^i_j \frac{\sqrt{1-2M/r_0}}{k_+ z} δ(r-r_0),\qquad T^r_r = 0.
\end{equation}
As a consistency check, this indeed satisfies \fref{eq:covd}. For this structure to end on the brane at $z=z_0$, $T^z_z$ must be further dressed up with a step function $\Theta(z-z_0)$,
\begin{equation}
    T^z_z = \frac{2\sqrt{1-2M/r_0}}{k_+²r_0 z²}\left(\frac{1-3M/2r_0}{\sqrt{1-2M/r_0}}-1\right) δ(r-r_0)\Theta(z-z_0),
\end{equation}
which gives
\begin{equation}
\begin{split}
    T^t_t + T^θ_θ + T^ϕ_ϕ = z ∂_z T^z_z + 4 T^z_z &= 
    \frac{4\sqrt{1-2M/r_0}}{k_+²r_0 z²}\left(\frac{1-3M/2r_0}{\sqrt{1-2M/r_0}}-1\right) δ(r-r_0)\Theta(z-z_0)\\
    &+\frac{2\sqrt{1-2M/r_0}}{k_+²r_0 z}\left(\frac{1-3M/2r_0}{\sqrt{1-2M/r_0}}-1\right) δ(r-r_0)δ(z-z_0).
\end{split}
\end{equation}
The delta function in $z$ arises from the derivative of the step function and induces matter on the dark bubble that exactly corresponds to the two dimensional shell of matter embedded in four dimensions.

\subsection{Uplifting downwards}\label{sec:rigid_shell:down}
There is no matter below the brane and the five dimensional metric is simply a vacuum solution to Einstein's equations with a negative cosmological constant. However, the five dimensional structure extending upwards will result in bending of the brane. One could, in principle, find a smooth global coordinate transformation that straightens out the bent brane (corresponding to a gauge choice in which the brane is located at a constant $z=z_0$), but we will not do this here, and will stick to the bent gauge. 

The spacetime below the brane (inside the dark bubble) deviates from empty AdS$_5$ in response to the bending of the brane and can be computed order by order in the bending, which is proportional to the mass induced on the dark bubble. Let us write this perturbed metric as
%
\begin{equation}\label{eq:perturbation}
\begin{split}
\D s²   &= \frac{\D z²}{k_-²z²} + k_-²z² η_{ab}\D x^a \D x^b + \overbrace{χ_{ab} \D x^a \D x^b}^{\rm{perturbation}} \\
&= \frac{\D z²}{k_-²z²} + k_-²z²\left[-\left(1+h_t(r,z)\right)\D t²+\left(1+h_r(r,z)\right)\D r²+\left(1+h_a(r,z)\right)r²\D \Omega_2²\right].
\end{split}
\end{equation}  
Choosing the perturbation to be traceless ($h_{ab}η^{ab}=0$), Einstein's equations at linear order in the perturbation give the following differential equation for the time component $h_{t}$:
\begin{equation}
∂_{rr}h_{t} + \frac{2}{r}∂_rh_{t}+5k^4z^3∂_zh_{t}+k_-^4z^4∂_{zz}h_{t}=0.
\end{equation}
To solve this equation, we can use the fact the time-time component of the metric gives the gravitational potential of the solution to linear order in the perturbation $χ/r$. It was shown in \cite{Banerjee:2020wix} that in vacuum, this is given by the Bessel function $K_2$ in momentum space. To get to position space, we need to superimpose the Bessel functions by integrating their Fourier transform over the shell of matter at $r=r_0$ (given by $\int\D \Omega_0$ below) which sources the solution. This gives 
\begin{equation}\label{eq:k2r0}
\begin{split}
h_t&=\int \D ^3 \vec{p} \int \D \Omega_0 e^{i\vec{p}\cdot \left(\vec{r}-\vec{r}_0\right)}K_2(p)
= 2π \int\limits_{-∞}^{∞} \D p \int\limits_{0}^{π}\D θ \sin θ e^{ipr\cos θ} p² K₂(p) \int \D \Omega_0 e^{ipr_0\cos θ_0}\\
&= 16π² \int\limits_{-∞}^{∞} \D p \frac{\sin pr}{pr} \frac{\sin pr_0}{pr_0} p² K₂(p)
= \frac{8π^3}{r r_0}\left[\frac{1+2k_-^4z²\left(r+r_0\right)²}{\sqrt{1+k_-^4z²\left(r+r_0\right)²}} - \frac{1+2k_-^4z²\left(r-r_0\right)²}{\sqrt{1+k_-^4z²\left(r-r_0\right)²}}\right].
\end{split}
\end{equation}
Einstein's equations further give a differential equation for the radial component of the perturbation in terms of the time component:
\begin{equation}
3 ∂_{z}h_{r} + r ∂_{rz}h_{r} + ∂_z h_{t}=0,
\end{equation}
which can be solved to give (we have dropped the constant $8\pi^3$ from $h_t$)
\begin{equation}\label{eq:k2r0-r}
h_{r} = \frac{1}{3r_0k_-^4z^3r^3}\left[ \left(1-k_-^4\left(2r-r_0\right)\left(r+r_0\right)z²\right) \sqrt{1+k_-^4z^2\left(r+r_0\right)^2} - \left(r_0 \mapsto -r_0\right) \right].
\end{equation}
The angular piece $h_a$ can be simply found from the tracelessness of the perturbation
$ h_t + h_r + 2h_a =0 $ to give
\begin{equation}\label{eq:k2r0-a}
h_a = \frac{1}{3r_0k_-^4z^3r^3} \left[ \frac{1+k_-^4z^2\left(2r²-rr_0+2r_0²\right)+k_-^8z^4\left(r-r_0\right)²\left(4r²+rr_0+r_0²\right)}{\sqrt{1+k_-^4z^2\left(r-r_0\right)^2}} - \left(r_0 \mapsto -r_0\right) \right].
\end{equation}
Note that in the limit $r_0 →0$, \ie, when the spherical shell of matter on the brane shrinks to a point source at $r=0$, the metric perturbation reduces to the result for a point source obtained in \cite{Banerjee:2020wix}. This is simply a consistency check, which ensures that the metric induced on the bent brane is, to leading order in  $M/r$, given by the Schwarzschild metric. In fact, the solution we have found here is exact up to the leading order in $M/r$ and all orders in $1/k_-²zr \sim R_\text{AdS}/r_p$, where $r_p$ is the proper radius.

\begin{figure}[h]
    \centering
    \begin{subfigure}{0.45\textwidth}
        \def\svgwidth{\linewidth}
        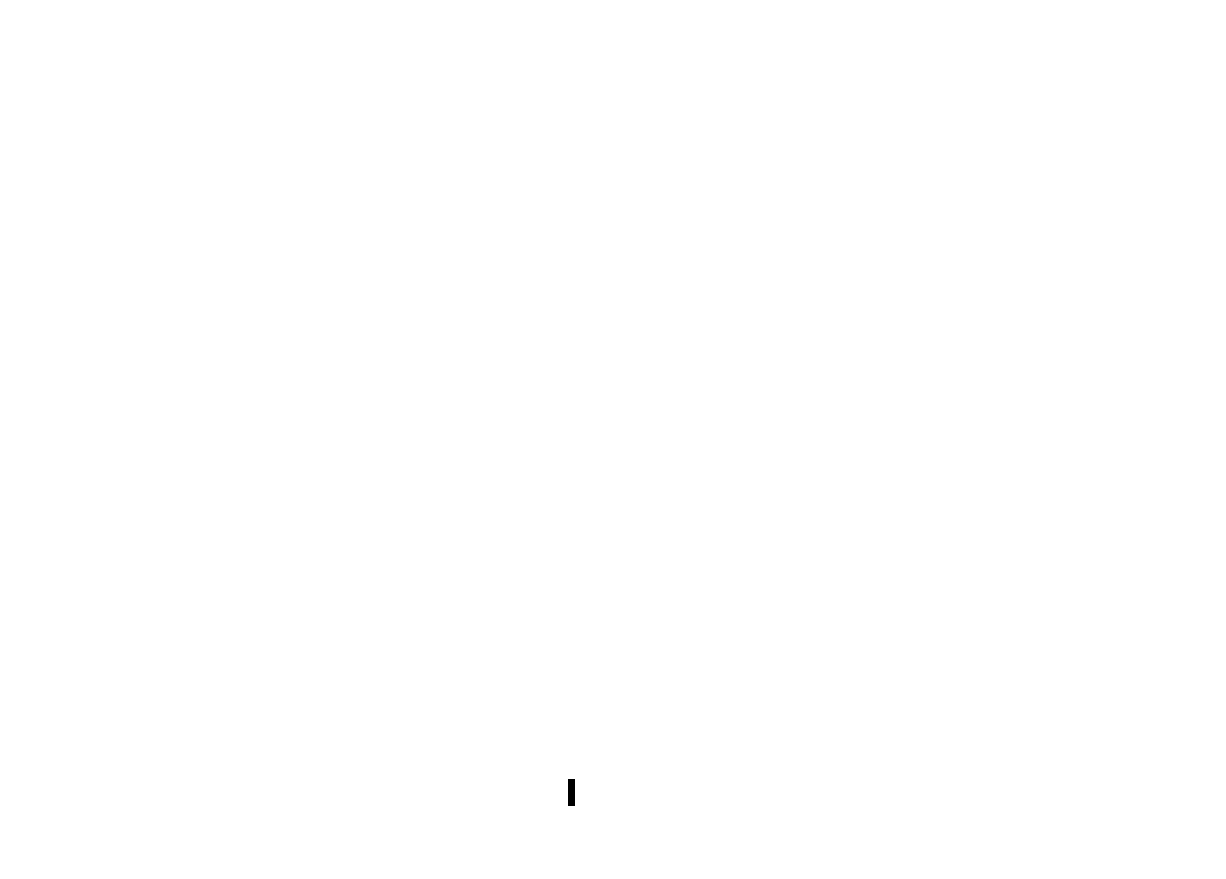
        \caption{}
    \end{subfigure}
    \hfill
    \begin{subfigure}{0.45\textwidth}
        \def\svgwidth{\linewidth}
        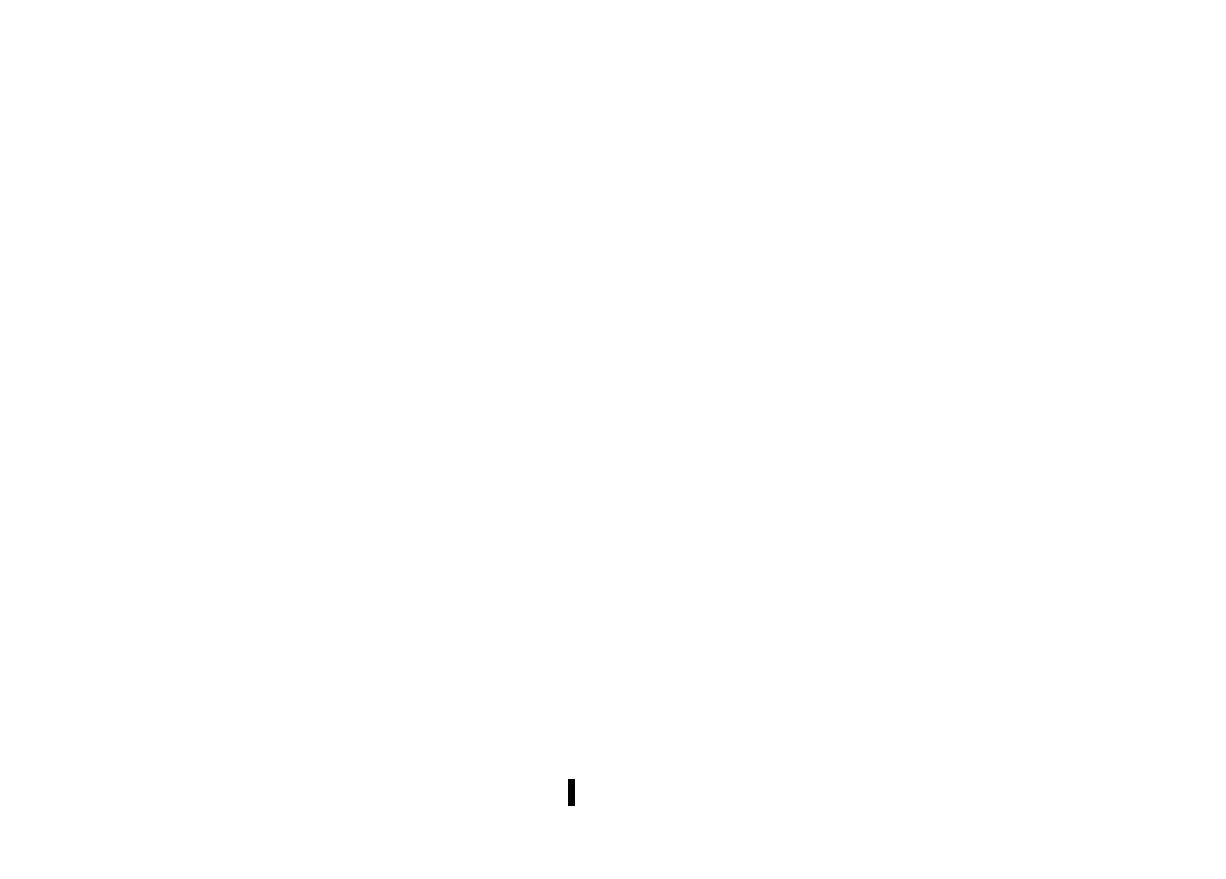
        \caption{}
    \end{subfigure}
    \caption{Gravitational potential below the brane as a function the coordinate distance $r$. This approaches a constant at small $r$ and falls off as $1/k_-r$ at large $r$, with a transition around $r=r_0$. The width of the transition is proportional to $~1/k_-²z$. This can be seen in the figures above where (b) is plotted for larger values of $k_-$ as compared to (a), for a fixed value of $z$, \eg, $z=2, r_0=1, k_-= 5$ for (a) and $k_-=15$ for (b).}
    \label{fig:plot_perturbation_ab}
\end{figure}

This is an interesting solution, which despite the parameter $r_0$, is a vacuum solution with no matter in the bulk. At small radius, the gravitational potential given by $h_t$ approaches a constant, 
which smoothly transitions into a $1/k_- r$  behavior far away as shown in \fref{fig:plot_perturbation_ab}. The width of the transition can be determined by taking the second derivative of the metric perturbation, which approaches a Gaussian of width $\sim 1/k_-²z$ close to $r_0$. Expressed in terms of proper length, the width of the shadow is given by $k_- z \times  1/k_-²z = k_-^{-1}$, which is the AdS-length. The transition becomes sharper as the AdS length becomes microscopic, but continues to be smooth.
Another way to think of this spacetime is that it is a solution to five dimensional Einstein's equations with a negative cosmological constant, with the boundary condition that it induces Schwarzschild geometry on a brane that bends in response to induced four dimensional matter.

We therefore have a metric below the brane that transitions smoothly from AdS at $r \ll r_0$ to that of a point source at $r \gg r_0$ with no matter localized at $r=r_0$. The transition, which takes place in vacuum, is made possible by five dimensional gravitational back-reaction of the spherical shell of matter and the corresponding structure on the outside.

This solution is also valid for a Randall-Sundrum braneworld, in which case reflection symmetry across the brane would imply the same vacuum solution on either side of the brane, without any material structure to support the spherical shell of matter in four dimensions. The brane would bend simply due to matter placed on it. 

\subsection{Sewing the bulk across the bubble wall}\label{sec:rigid_shell:updown}

To summarize, we have generated a structure that extends outwards from the bubble wall in the fifth dimension and its endpoint on the dark bubble induces a macroscopic shell of four dimensional matter. We have performed the analysis in the limit where the shell of matter in the four dimensional braneworld is much bigger than its Schwarzschild radius $r_0 \gg 2M$, but we expect the shadow shell to be the structure that persists to all orders. The matter source on the brane causes it to bend, which in turn sources the geometry below the brane (inside of the dark bubble), given in equations \eqref{eq:k2r0}, \eqref{eq:k2r0-r} and \eqref{eq:k2r0-a}. Pictorially one can imagine how the two pieces in \fref{fig:structure_above_below:a} and  \fref{fig:structure_above_below:b} can be stitched together. Whether we come from above or from below, the induced metric on the dark bubble will be Schwarzschild (at linear order) and all junction conditions will be satisfied. 

It is worth stressing here again that the two different choices of gauges on the two sides of the dark bubble make the analysis simpler in this construction. While above the brane, where there is a source, we specifically worked with the straight gauge, it was convenient to choose bent gauge in the construction of the five dimensional bulk spacetime below the brane, where there is no source. These two different gauges are useful to have, for a fully consistent five dimensional picture on either side of the brane while producing the same effective four dimensional gravity on the brane. This makes the analysis of junction conditions across the brane extremely interesting. In \cite{Banerjee:2020wix} we discussed the computation of the four dimensional gravitational propagator in momentum space using straight gauge, and argued that in order to achieve the correct behavior we needed modes in the form of the Bessel function $K_2$ on one side while a specific combination of  Bessel functions $K_2$ and $I_2$ on the other side. This combination was responsible for a purely non-normalizable behavior of the modes on our dark bubble. Furthermore, we also mentioned that in the case of a bent gauge, to reproduce the same four dimensional gravity on the braneworld, we need only $K_2$ on both sides of the brane.

Note that while considering the metric below the brane in the present work, we actually only considered pure $K_2$. Therefore our present construction relying on two different choices of gauges inside and outside the dark bubble, is interesting from the perspective of mode mixing. It is possible that one could have a bit of $I_2$ modifying higher order corrections of the metric outside of the dark bubble, and possibly on the braneworld as well. However, we leave a detailed analysis of this for a future work.

\section{Black holes on dark bubbles}\label{sec:blackholes}

If we want to consider black holes on our dark bubbles, we must go beyond linearized gravity. In the preceding section, we discussed a solution to leading order in $M/r$ and to all orders in $1/k²zr \sim R_\text{AdS}/r_p $. The question is whether it is possible to find this solution to all orders in $M/r$. We will try to answer this question in this section.

Let us first discuss the global properties of the spacetime below the brane. The first order expression is valid as long as $M/r$ is small. Is it possible to move away from the brane on the inside (small $z$) and pass through $r=0$ using just the first order corrections? To see whether this is possible, let us note that the proper Schwarzschild radius on the brane is $\sim kMz_0$. The Schwarzschild radius is much larger than the AdS$_5$ radius, so $k²Mz_0 \gg 1$. The radius of the dark bubble (determining the curvature of the universe) is even larger, so $kz_0 \gg k²Mz_0 \gg 1$, which implies that $M \ll 1/k$. The proper Schwarzschild radius away from the brane (in units of AdS$_5$ radius) at some $z$ is $kMz$. It turns out that the first order corrections obtained in \cite{Banerjee:2020wix} are finite when $r=0$, and proportional to $k^2Mz$. For $k²Mz \ll 1$, so that it is enough to consider first order results in $M/r$, we need that $z$ is small enough. Hence, the metric goes to the AdS$_5$ metric for small enough $z$, which is to be expected since this is a mode designed to vanish as $z$ goes to zero. Therefore, if the spacetime below the brane were to have a horizon, it will be capped off and contained above some finite and small $z$.

It is furthermore possible to systematically calculate corrections to all orders in $1/k^2 zr$, order by order in $\chi/r$. Note that the orders are coupled and care is needed when truncating the expansions. This can be done in Gauss normal coordinates, where the metric can be written as
\begin{equation}
	\D s² = \frac{\D z²}{k²z²} + k²z²\left(-g_{tt}\D t²+g_{rr}\D r²+g_{θθ}r²\D \Omega_2²\right).
\end{equation}
To linear order in $χ/r$ and quadratic order in $1/k²zr$, Einstein's equations with a negative cosmological constant give
\begin{equation}\label{eq:exactchizeroth}
	g_{tt} = 1-\frac{4χ}{r},\quad 
	g_{rr} = 1+\frac{χ}{r}\left(2-\frac{1}{k^4r²z²}\right),\quad
	g_{θθ} = 1+\frac{χ}{r}\left(1+\frac{1}{2k^4r²z²}\right).
\end{equation}
At leading (zeroth) order in $1/k²rz$, this leads to the familiar expression from \cite{Banerjee:2020wix}
\begin{equation}
	\D s² = \frac{\D z²}{k²z²} + k²z²\left[-\left(1-\frac{4χ}{r}\right)\D t²+\left(1+\frac{2χ}{r}\right)\D r²+\left(1+\frac{χ}{r}\right)r²\D \Omega_2²\right].
\end{equation} 
To second order in $1/k²rz$ and second order in $χ/r$, we get
\begin{equation}
\begin{split}
	g_{tt} &= 1-\frac{4χ}{r}
		+\frac{χ²}{r²}\left(\frac{45}{8}-c_1+\frac{7}{8k^4r²z²}\right),\\
	g_{rr} &= 1+\frac{χ}{r}\left(2-\frac{1}{k^4r²z²}\right)
		+\frac{χ²}{r²}\left(\frac{5}{4}-2c_1-c_2+\frac{2+3c_1}{k^4r²z²}\right),\\
	g_{θθ} &= 1+\frac{χ}{r}\left(1+\frac{1}{2k^4r²z²}\right)
        +\frac{χ²}{r²}\left(c_2+\frac{c_1}{k^4r²z²}\right),
\end{split}
\end{equation}
with a number of free parameters $c_i$. We can continue to solve Einstein's equations perturbatively to higher orders in $χ/r$ and $1/k²zr$, with more free parameters. However, in practice, we are mainly interested in the metric to all orders in $\chi/r$, and leading order in $1/k^2 rz$. 
To find this, it is convenient to choose a gauge where the determinant of the metric is independent of the perturbations at leading order in $1/k^2 rz$. Furthermore, we can impose that there are no corrections beyond linear order in the angular part of the metric. Remarkably, we then find that the perturbative solution can be re-summed at zeroth order in $1/k^2 rz$ to give a metric that is an {\it exact} solution to all orders in $χ/r$
\begin{equation}\label{eq:exactzeroth}
    \D s_5² =  k²z²\left[-\frac{1-2χ/r}{\left(1+χ/r\right)^2}\D t²+\frac{1}{\left(1-2χ/r\right)}\D r²+\left(1+\frac{χ}{r}\right)r²\D \Omega_2²\right] + \frac{\D z²}{k²z²}.
\end{equation}
A brane embedded in this geometry at
\begin{equation}\label{eq:embedding_linear}
    z(r)=z_0 \sqrt{1+\frac{\chi}{r}},
\end{equation}
has an induced metric that is \emph{exactly} Schwarzschild (to all orders in $χ/r$ and zeroth order in $1/k²rz$)
\begin{equation}\label{eq:schw4}
	\D s_4² = k²z_0²\left[-\left(1-\frac{2M}{ρ}\right)\D t² + \frac{\D r²}{1-2M/ρ}+ρ²\D \Omega_2²\right],
\end{equation}
where $\rho=r+\chi$ is the new radial coordinate. We also note that $M=3χ/2$ consistent with the results obtained in \cite{Banerjee:2020wix, Banerjee:2020wov}. Note that the contribution from $\D z^2/\left(k²z²\right)$ is sub-leading and does not contribute at this order. We can also re-sum the perturbation series at the next order (linear) in $1/k^2rz$, to obtain an exact solution to all orders in $χ/r$ given by
\begin{equation}\label{eq:exactchisecond}
\begin{split}
	g_{tt} &= 
	\frac{1-2χ/r}{\left(1+χ/r\right)^2}
	+\frac{1}{8k⁴r²z²}\cdot\frac{χ²}{r²}\cdot\frac{\left(1-2χ/r\right)\left(7-2χ/r\right)}{\left(1+χ/r\right)^4},\\
	g_{rr} &= 
	\frac{1}{\left(1-2χ/r\right)}
	-\frac{1}{8k⁴r²z²}\cdot\frac{χ}{r}\cdot\frac{8-17χ/r-10χ²/r²}{\left(1-2χ/r\right)\left(1+χ/r\right)²},\\
	g_{θθ} &=
	\left(1+\frac{χ}{r}\right)
	+ \frac{1}{8k⁴r²z²}\cdot\frac{χ}{r}\cdot \frac{\left(1-2χ/r\right)\left(4+χ/r\right)}{\left(1+χ/r\right)},
\end{split}
\end{equation}
which gives an exact (in $χ/r$) embedding of the Schwarzschild brane at
\begin{equation}
	z(r) = z_0 \sqrt{1+\frac{χ}{r}}
    \left(1-\frac{1}{16k⁴r²z_0²}\cdot\frac{χ²}{r²}\cdot\frac{1-2χ/r}{\left(1+χ/r\right)^3} \right).
\end{equation}
The induced metric is Ricci flat and represents a Schwarzschild geometry to quadratic order in $1/k^2rz_0$ and all orders in $χ/r$. A coordinate transformation can be performed to change  to Schwarzschild coordinates as in \fref{eq:schw4}, but we will not do that here. 

\begin{figure}
    \centering
    \def\svgwidth{0.6\linewidth}
    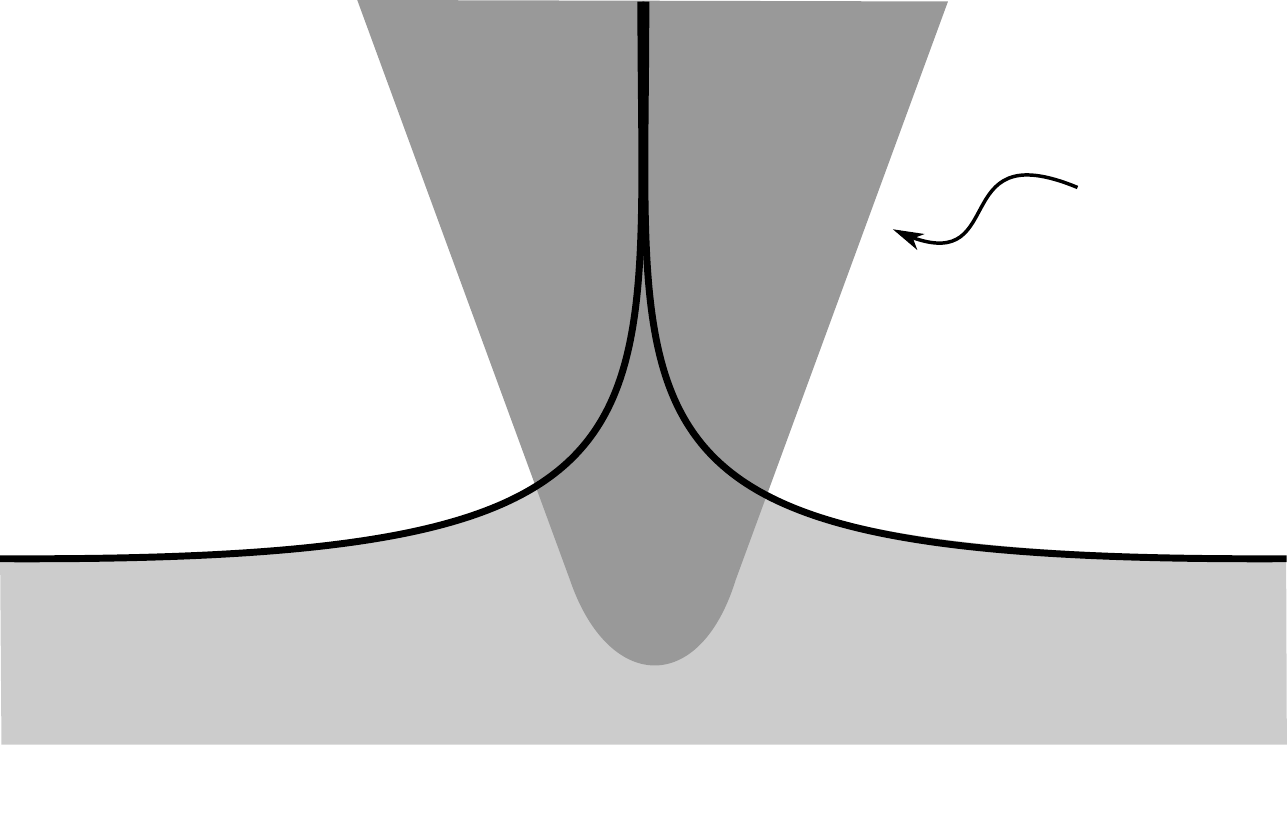
    \caption{A dark  bubble with a black hole with its interior shown in dark gray. The horizon extends below the brane, into the dark bubble and is smoothly capped off at small $z$.}
\label{fig:blackhole}
\end{figure}

The metric obtained in \fref{eq:exactchisecond} is quite remarkable in that it has a coordinate singularity at $r=2χ$, but has no curvature singularities. Moreover, there is no source below the brane at $r=0$. The horizon at $r=2χ$ is therefore a \emph{shadow} horizon and arises due to the bending of the brane that gets capped off at a finite but small $z$. This is shown in \fref{fig:blackhole}. This is a remarkable geometry, and as discussed in \fref{sec:rigid_shell}, can represent black holes on our dark bubbles as well as braneworld black holes for a Randall-Sundrum geometry. This could be relevant for the old problem of finding an explicit metric of a four dimensional braneworld black hole mentioned in, e.g., \cite{Emparan:1999wa}, and generalizing recent results for three dimensional BTZ black holes in, e.g., \cite{Emparan:2020znc}. Let us reiterate that for a Randall-Sundrum braneworld, the brane can bend in response to matter on the brane and this is the geometry on either side of the brane.

Note that in order to obtain the full metric on both sides, one needs to re-sum to all orders in $1/k²zr$. Since we have not been able to do that here, we can only reliably talk about the metric when this quantity is small enough that the above result at linear order makes sense. In the above, we have a reliable picture of the horizon of the black structure that supports the black hole on the dark bubble from outside at leading order in $1/k²zr$. To study the nature of the horizon inside of the dark bubble, one needs to move into the bulk, away from the bubble wall. For this, one needs the full metric re-summed to all orders in $1/k²zr$. However, based on the discussion in \fref{sec:rigid_shell}, we would expect there to be a shadow shell with a smoothly capped geometry inside the bubble. Based on this intuition, let us speculate the nature of the full geometry below.

The metric above, combined with the shadow shell of \fref{sec:rigid_shell} on the inside, leads to the intriguing geometry sketched in \fref{fig:blackhole}. The brane only bends gently before entering through the horizon, with $z$ increasing by a factor of only $\sqrt{3/2}$. Inside the horizon, however, the metric diverges towards $r=0$ where the singularity sits at $z \rightarrow \infty$. Note how this is consistent with the fact that our metric was constructed, in the Poincaré patch corresponding to the limit of a large dark bubble, under the assumption that there were no sources. The horizon that we find requires a singularity but it sits at infinity. What would an observer on the dark bubble observe entering through the horizon? Inside a Schwarzschild black hole, using Schwarzschild coordinates, one finds a time dependent geometry that contracts towards the future singularity in a finite time. $r$ is time, while $t$ is an infinite space-like coordinate. The same story plays out regardless of when one enters the black hole. The only difference is where in the spatial coordinate $t$ that one enters. The dark bubble adds another twist to the story. While the observations within the four dimensional world are identical, there is also a five dimensional interpretation. What happens when you enter through the horizon is that you step into an elevator that brings you up in $z$ so that you hit the singularity at infinity in a finite time. At whatever time that you enter, the elevator will be ready to bring you along on its journey. One might wonder how the time-like brane fits with a space-like singularity, but it works just in the same way as with space time outside and inside of a black hole horizon. Inside of the horizon, it is the radial direction that is time. This direction ends when you reach the space-like singularity at $z=\infty$, which is spanned by the spatial direction that was time outside of the horizon. Depending on when you enter into the black hole you will hit the singularity at different points along its length.
\section{Discussion and Outlook}

In this article we have demonstrated that our model proposed in \cite{Banerjee:2018qey}, which can incorporate dark energy into string theory, is also capable of describing non-trivial gravitational phenomena such as black holes. In the process, we have also discussed another very interesting toy example of a thin shell of matter in four dimensions and its evolution in the fifth dimension leading to two completely different bulk extensions above and below our dark bubble. The two different bulk extensions were necessitated by the fact that the five dimensional spacetime above the dark bubble is sourced by physical strings attached to it, while there is no source below the dark bubble. Our challenge was to combine these two five dimensional pictures in such a way that they reproduce the same gravitational description on our dark bubble. In particular, if we look from above, the end points of the strings result in an effective Schwarzschild geometry on the dark bubble. Therefore, it was natural to expect that the same geometry should be induced on the brane also from the five dimensional bulk geometry inside the dark bubble. In this work we showed that this can actually be achieved by imposing different boundary conditions from either side of the dark bubble.

We then proceed further by adding perturbative corrections  in the form of expansions in $M/r$ and $R_\text{AdS}/r_p$ to the metric outside the dark bubble. We are able to obtain a solution, exact in all orders in $M/r$  and at linear order in $R_\text{AdS}/r_p$, which induces a Schwarzschild metric on our dark bubble. This metric has a horizon and a singularity that sits at $r→0$ and $z→∞$. We also find the solution to the next perturbative order in $R_\text{AdS}/r_p$, but are unable to re-sum the corrections to find the find the full bulk metric. This prevents us from writing down the solution inside the bubble, away from the dark bubble. However, since the spacetime inside the bubble is empty AdS$_5$, we expect this vacuum solution to be of the same form as the matter shell that we have obtained. Based on this intuition, we propose in \fref{fig:blackhole}, the full solution representing a black hole on the dark bubble.
Clearly, there will be corrections when the scale of the horizon approaches $R_\text{AdS}$, or when one is too close to the singularity. Such corrections will appear in the bulk and will affect the four dimensional expressions as well. Similarly, the expansion that we did to obtain the Friedmann cosmology will receive corrections when the energy densities are too high. In addition to these kinds of corrections, which in principle are calculable (if one could solve the Einstein equations to all orders in $M/r$ and $R_\text{AdS}/r_p$ at the same time), there can also be loop contributions that we have not taken into account.

Let us briefly comment on some braneworld black hole solutions and their possible relation to the solution that we have found above.\footnote{We would like to thank the referee for pointing out these interesting results and references.} While the CHR solution \cite{Chamblin:1999by} describes a black string with an induced Schwarzschild metric on the bubble wall, it is known to suffer from horizon shredding due to the Gregory-Laflamme instability \cite{Gregory:1993vy}. A very promising solution is that constructed by Plebanski-Demianski solution \cite{PLEBANSKI197698}, which is the AdS extension of the C-metric \cite{PhysRevD.2.1359} and is referred to as the AdS-C metric. This is a four dimensional metric that describes accelerated black holes with conical deficit at one or both poles. These are interpreted as cosmic strings pulling on the black holes and accelerating them. This metric has been studied extensively in the context of Randall-Sundrum braneworlds \eg, in \cite{Emparan:1999wa,Emparan:1999fd}. Although attempts have been made to construct analogous solutions in five dimensions, to the best of our knowledge, there are no known analytic five dimensional AdS-C metrics. Since the four dimensional AdS-C metric looks like an accelerating black hole being pulled by a radially stretched cosmic string, it is indeed a candidate for a black hole on the dark bubble. This is a very interesting and relevant solution and we would like to construct a four dimensional analog of our present solution to compare it with the results obtained from the AdS-C metric. 
	
Some other solutions of interest include black droplets and black funnels \cite{Hubeny:2009ru}, which describe black holes on the brane with their horizon either disconnected (black droplet) or connected smoothly (black funnel) to the Poincaré horizon in the bulk. The black droplet metric looks qualitatively similar to the metric inside our dark bubble and once we have the re-summed metric on the inside of the brane, they should indeed be compared. However, as with the AdS-C metric, these metrics are analytically known only for AdS$_4$ with only numerical results available in five dimensions \cite{Figueras:2011va,Figueras:2011gd}.

Although it is immensely encouraging to find such a black hole like solution, it is far from clear whether such a configuration yields a realistic outcome of gravitational collapse. While the solution formally exists, as we have shown, we believe that there are new possibilities that distinguish our model from Randall-Sundrum braneworlds. In particular, there is the intriguing possibility of realizing the black shell alternative to a black hole proposed in \cite{Danielsson:2017riq,Danielsson:2017pvl}. We will address this question in an upcoming work.

\section*{Acknowledgments}
SG would like to thank Uppsala University for their kind hospitality where this work was completed. The work of SB is supported by the Alexander von Humboldt postdoctoral fellowship. SG is partially supported by the INFN and the MIUR-PRIN contract 2017CC72MK003AZ. We would also like to thank the anonymous referee for her/his invaluable suggestions that have immensely helped to improve this article.

    \small
    \bibliography{references}
    \bibliographystyle{utphysmodb}
\end{document}

%% file: figures/straight_outside.pdf_tex
\begingroup%
  \makeatletter%
  \providecommand\color[2][]{%
    \errmessage{(Inkscape) Color is used for the text in Inkscape, but the package 'color.sty' is not loaded}%
    \renewcommand\color[2][]{}%
  }%
  \providecommand\transparent[1]{%
    \errmessage{(Inkscape) Transparency is used (non-zero) for the text in Inkscape, but the package 'transparent.sty' is not loaded}%
    \renewcommand\transparent[1]{}%
  }%
  \providecommand\rotatebox[2]{#2}%
  \newcommand*\fsize{\dimexpr\f@size pt\relax}%
  \newcommand*\lineheight[1]{\fontsize{\fsize}{#1\fsize}\selectfont}%
  \ifx\svgwidth\undefined%
    \setlength{\unitlength}{191.44915964bp}%
    \ifx\svgscale\undefined%
      \relax%
    \else%
      \setlength{\unitlength}{\unitlength * \real{\svgscale}}%
    \fi%
  \else%
    \setlength{\unitlength}{\svgwidth}%
  \fi%
  \global\let\svgwidth\undefined%
  \global\let\svgscale\undefined%
  \makeatother%
  \begin{picture}(1,0.55602129)%
    \lineheight{1}%
    \setlength\tabcolsep{0pt}%
    \put(0,0){\includegraphics[width=\unitlength,page=1]{figures/straight_outside.pdf}}%
    \put(0.43441957,0.36487561){\color[rgb]{0,0,0}\makebox(0,0)[lt]{\lineheight{1.25000012}\smash{\begin{tabular}[t]{l}AdS$_5$\end{tabular}}}}%
    \put(0.03875318,0.3599022){\color[rgb]{0,0,0}\makebox(0,0)[lt]{\lineheight{1.25000012}\smash{\begin{tabular}[t]{l}CHR\end{tabular}}}}%
    \put(0.83162716,0.304246){\color[rgb]{0,0,0}\makebox(0,0)[lt]{\lineheight{1.25000012}\smash{\begin{tabular}[t]{l}$k_+$\end{tabular}}}}%
    \put(0.8384369,0.15920043){\color[rgb]{0,0,0}\makebox(0,0)[lt]{\lineheight{1.25000012}\smash{\begin{tabular}[t]{l}$k_-$\end{tabular}}}}%
    \put(0,0){\includegraphics[width=\unitlength,page=2]{straight_outside.pdf}}%
  \end{picture}%
\endgroup%

%% file: figures/bent_inside.pdf_tex
\begingroup%
  \makeatletter%
  \providecommand\color[2][]{%
    \errmessage{(Inkscape) Color is used for the text in Inkscape, but the package 'color.sty' is not loaded}%
    \renewcommand\color[2][]{}%
  }%
  \providecommand\transparent[1]{%
    \errmessage{(Inkscape) Transparency is used (non-zero) for the text in Inkscape, but the package 'transparent.sty' is not loaded}%
    \renewcommand\transparent[1]{}%
  }%
  \providecommand\rotatebox[2]{#2}%
  \newcommand*\fsize{\dimexpr\f@size pt\relax}%
  \newcommand*\lineheight[1]{\fontsize{\fsize}{#1\fsize}\selectfont}%
  \ifx\svgwidth\undefined%
    \setlength{\unitlength}{191.09364424bp}%
    \ifx\svgscale\undefined%
      \relax%
    \else%
      \setlength{\unitlength}{\unitlength * \real{\svgscale}}%
    \fi%
  \else%
    \setlength{\unitlength}{\svgwidth}%
  \fi%
  \global\let\svgwidth\undefined%
  \global\let\svgscale\undefined%
  \makeatother%
  \begin{picture}(1,0.31344493)%
    \lineheight{1}%
    \setlength\tabcolsep{0pt}%
    \put(0,0){\includegraphics[width=\unitlength,page=1]{figures/bent_inside.pdf}}%
    \put(0.82261559,0.27728895){\color[rgb]{0,0,0}\makebox(0,0)[lt]{\lineheight{1.25000012}\smash{\begin{tabular}[t]{l}$k_+$\end{tabular}}}}%
    \put(0.82943799,0.13197353){\color[rgb]{0,0,0}\makebox(0,0)[lt]{\lineheight{1.25000012}\smash{\begin{tabular}[t]{l}$k_-$\end{tabular}}}}%
    \put(0,0){\includegraphics[width=\unitlength,page=2]{bent_inside.pdf}}%
  \end{picture}%
\endgroup%

%% file: figures/fig1a.pdf_tex
\begingroup%
  \makeatletter%
  \providecommand\color[2][]{%
    \errmessage{(Inkscape) Color is used for the text in Inkscape, but the package 'color.sty' is not loaded}%
    \renewcommand\color[2][]{}%
  }%
  \providecommand\transparent[1]{%
    \errmessage{(Inkscape) Transparency is used (non-zero) for the text in Inkscape, but the package 'transparent.sty' is not loaded}%
    \renewcommand\transparent[1]{}%
  }%
  \providecommand\rotatebox[2]{#2}%
  \newcommand*\fsize{\dimexpr\f@size pt\relax}%
  \newcommand*\lineheight[1]{\fontsize{\fsize}{#1\fsize}\selectfont}%
  \ifx\svgwidth\undefined%
    \setlength{\unitlength}{352.89160682bp}%
    \ifx\svgscale\undefined%
      \relax%
    \else%
      \setlength{\unitlength}{\unitlength * \real{\svgscale}}%
    \fi%
  \else%
    \setlength{\unitlength}{\svgwidth}%
  \fi%
  \global\let\svgwidth\undefined%
  \global\let\svgscale\undefined%
  \makeatother%
  \begin{picture}(1,0.71679844)%
    \lineheight{1}%
    \setlength\tabcolsep{0pt}%
    \put(0,0){\includegraphics[width=\unitlength,page=1]{figures/fig1a.pdf}}%
    \put(-0.001439,0.6961681){\color[rgb]{0,0,0}\makebox(0,0)[lt]{\lineheight{1.25}\smash{\begin{tabular}[t]{l}$h_{t}$\end{tabular}}}}%
    \put(0.45975078,0.00561765){\color[rgb]{0,0,0}\makebox(0,0)[lt]{\lineheight{1.25}\smash{\begin{tabular}[t]{l}$r_0$\end{tabular}}}}%
    \put(0.95919593,0.00332078){\color[rgb]{0,0,0}\makebox(0,0)[lt]{\lineheight{1.25}\smash{\begin{tabular}[t]{l}$r$\end{tabular}}}}%
    \put(0,0){\includegraphics[width=\unitlength,page=2]{fig1a.pdf}}%
    \put(1.06971144,0.14784107){\color[rgb]{0,0,0}\makebox(0,0)[lt]{\begin{minipage}{1.1264082\unitlength}\raggedright \end{minipage}}}%
  \end{picture}%
\endgroup%

%% file: figures/fig1b.pdf_tex
\begingroup%
  \makeatletter%
  \providecommand\color[2][]{%
    \errmessage{(Inkscape) Color is used for the text in Inkscape, but the package 'color.sty' is not loaded}%
    \renewcommand\color[2][]{}%
  }%
  \providecommand\transparent[1]{%
    \errmessage{(Inkscape) Transparency is used (non-zero) for the text in Inkscape, but the package 'transparent.sty' is not loaded}%
    \renewcommand\transparent[1]{}%
  }%
  \providecommand\rotatebox[2]{#2}%
  \newcommand*\fsize{\dimexpr\f@size pt\relax}%
  \newcommand*\lineheight[1]{\fontsize{\fsize}{#1\fsize}\selectfont}%
  \ifx\svgwidth\undefined%
    \setlength{\unitlength}{352.89160682bp}%
    \ifx\svgscale\undefined%
      \relax%
    \else%
      \setlength{\unitlength}{\unitlength * \real{\svgscale}}%
    \fi%
  \else%
    \setlength{\unitlength}{\svgwidth}%
  \fi%
  \global\let\svgwidth\undefined%
  \global\let\svgscale\undefined%
  \makeatother%
  \begin{picture}(1,0.71679844)%
    \lineheight{1}%
    \setlength\tabcolsep{0pt}%
    \put(0,0){\includegraphics[width=\unitlength,page=1]{figures/fig1b.pdf}}%
    \put(-0.001439,0.6961681){\color[rgb]{0,0,0}\makebox(0,0)[lt]{\lineheight{1.25}\smash{\begin{tabular}[t]{l}$h_{t}$\end{tabular}}}}%
    \put(0.45975078,0.00561765){\color[rgb]{0,0,0}\makebox(0,0)[lt]{\lineheight{1.25}\smash{\begin{tabular}[t]{l}$r_0$\end{tabular}}}}%
    \put(0.95919593,0.00332078){\color[rgb]{0,0,0}\makebox(0,0)[lt]{\lineheight{1.25}\smash{\begin{tabular}[t]{l}$r$\end{tabular}}}}%
    \put(0,0){\includegraphics[width=\unitlength,page=2]{fig1b.pdf}}%
  \end{picture}%
\endgroup%

%% file: figures/horizon.pdf_tex
\begingroup%
  \makeatletter%
  \providecommand\color[2][]{%
    \errmessage{(Inkscape) Color is used for the text in Inkscape, but the package 'color.sty' is not loaded}%
    \renewcommand\color[2][]{}%
  }%
  \providecommand\transparent[1]{%
    \errmessage{(Inkscape) Transparency is used (non-zero) for the text in Inkscape, but the package 'transparent.sty' is not loaded}%
    \renewcommand\transparent[1]{}%
  }%
  \providecommand\rotatebox[2]{#2}%
  \newcommand*\fsize{\dimexpr\f@size pt\relax}%
  \newcommand*\lineheight[1]{\fontsize{\fsize}{#1\fsize}\selectfont}%
  \ifx\svgwidth\undefined%
    \setlength{\unitlength}{370.61503588bp}%
    \ifx\svgscale\undefined%
      \relax%
    \else%
      \setlength{\unitlength}{\unitlength * \real{\svgscale}}%
    \fi%
  \else%
    \setlength{\unitlength}{\svgwidth}%
  \fi%
  \global\let\svgwidth\undefined%
  \global\let\svgscale\undefined%
  \makeatother%
  \begin{picture}(1,0.638533)%
    \lineheight{1}%
    \setlength\tabcolsep{0pt}%
    \put(0,0){\includegraphics[width=\unitlength,page=1]{figures/horizon.pdf}}%
    \put(0.84840896,0.47948947){\color[rgb]{0,0,0}\makebox(0,0)[lt]{\lineheight{1.25}\smash{\begin{tabular}[t]{l}horizon\end{tabular}}}}%
    \put(0.03902083,0.28052617){\color[rgb]{0,0,0}\makebox(0,0)[lt]{\lineheight{1.25}\smash{\begin{tabular}[t]{l}$k_+$\end{tabular}}}}%
    \put(0.04253853,0.13570254){\color[rgb]{0,0,0}\makebox(0,0)[lt]{\lineheight{1.25}\smash{\begin{tabular}[t]{l}$k_-$\end{tabular}}}}%
  \end{picture}%
\endgroup%